\documentstyle[aps,preprint,prb]{revtex}

\begin{document}

\title{Quantum Monte Carlo Study of Hole Binding and Pairing
Correlations in the Three-Band Hubbard Model} 
\author {M.\ Guerrero and J. \ E. \ Gubernatis}
\address{
Theoretical Division, Los Alamos National Laboratory, Los Alamos, NM 87545}
\author{Shiwei Zhang } 
\address{Department of Physics and Department of Applied Science\\
 College of William and Mary, Williamsburg, VA 23187}

\date{ \today}

\maketitle
\begin{abstract}
We simulated the 3-band Hubbard model using the Constrained Path 
Monte Carlo (CPMC) method
in search for a possible superconducting ground state. The CPMC is a ground 
state method which is free of the exponential scaling of computing time with 
system size. We calculated
the binding energy of a pair of  holes for systems up to $6 \times 4$ unit 
cells.
We also studied  the pairing correlation functions versus distance 
for both the d-wave and extended s-wave channels in systems up to $6 \times 6$.
We found that
holes bind for a wide range of parameters and that
the binding increased as the system size is increased. However, the pairing 
correlation functions decay quickly with distance.
 For
the extended s channel, we found that as the Coulomb interaction $U_d$
on the Cu sites is increased, the long-range part of the correlation
functions is suppressed and fluctuates around zero. For the $d_{x^2 -
y^2}$ channel, we found that the correlations decay rapidly with
distance towards a small positive value. However, this value becomes
smaller as the interaction $U_d$ or the system size is increased.

\vspace{1cm}
\end{abstract}

\section{Introduction}
The discovery of high temperature superconductivity in copper oxide
materials generated considerable effort to find simple electronic
models that exhibit a superconducting phase. Since the mechanism for
superconductivity is believed to lie within the CuO$_2$ planes, 
 a variety of two-dimensional models have been
proposed, the three principal ones being the (one-band) Hubbard,
three-band Hubbard, and t-J models. Unfortunately, conclusive evidence
for a superconducting phase has not been found in any of these models.
\cite{elbio} 

In this paper, we will report the results of a quantum Monte Carlo
(QMC) study of the three-band Hubbard model. Of the three models, it
has been less intensively studied, partially because of the 
general belief that its low energy
excitation spectrum is similar to the other two. Indeed in the strong
coupling limit, both the one-band and three-band Hubbard models have the
t-J model as an approximate limit. However, there still remains some
controversy on whether one-band models, like the Hubbard and t-J
models, are adequate to describe the low energy physical properties of
the cuprate superconductors. One of our objectives is to study the
possible existence of superconductivity in the three-band model in
regions where model parameters are physical as opposed to regions where
asymptotic models are clearly more appropriate.

The most solid information about possible
superconductivity in the one-band Hubbard model has come from a series
of QMC calculations.  For instance, using a finite temperature QMC
method, White et al. \cite{white1} studied possible superconductivity 
in the Hubbard model. 
They found an attractive effective pairing interaction in
the $d_{x^2 - y^2}$ and extended s-wave channels.  Moreo and Scalapino
\cite{moreo} subsequently found pairing correlations but also found that they
did not increase when the lattice size was increased. Their result, suggesting
the absence of long range order, was consistent with an earlier QMC
study by Imada and Hatsugai.\cite{imada}

The results of almost all QMC calculations of the Hubbard model
however, are limited by the fermion sign problem to high temperatures
and small system sizes. These limitations had left open the
possibility that superconductivity still lurked at larger systems and
lower temperatures.  Recently, a new QMC method, the Constrained Path
Monte Carlo (CPMC) method,\cite{shiwei1} was developed to get around
the sign problem.  Using this method, Zhang et al. \cite{shiwei2}
calculated ground-state pairing correlation functions for the Hubbard
model as a function of the distance and found that as the system size
or the interaction strength is increased the magnitude of the
long-range part of the pairing correlation functions vanished for both
the $d_{x^2 - y^2}$ and extended s-wave channels. Although this method
produces a variational solution, its results, together with the null
results from previous QMC studies, are very discouraging for
finding superconductivity in the one-band  Hubbard model.

Past numerical work on the three-band model,  has been less definitive about
the existence of superconductivity. The work has mainly consisted of
exact diagonalization computations \cite{ogata}$^{-}$\cite{stephan}, 
where calculations of hole binding
energies was emphasized, and QMC computations \cite{scalettar1}$^{-}$ 
\cite{kuroki} where magnetic and
pairing calculations were emphasized.  The exact diagonalization
studies unequivocally established that holes can bind. In a number of
cases binding only occurred for unphysical parameter ranges, and
because of small systems sizes, the existence and symmetry of the
pairing was difficult to access. The QMC studies, probably even more
limited by the sign problem than those for the one-band model, 
 established the existence of an extended s-wave and $d_{x^2-y^2}$ 
wave attractive pairing interaction. 
However, none of these QMC studies reported the behavior of the pairing 
correlation functions at large distances, which is the relevant quantity
to establish the existence of long range order.

In the work reported here, we removed the limitations caused by the
sign problem by applying the CPMC method to the three-band Hubbard
model. We computed the binding energy for holes near
half-filling for systems up to $6 \times 4$ unit cells. We found a wide range 
of physically relevant
parameters for which the holes bind. As will be reported in more
detail, we found that larger systems in fact make hole binding easier. We also
find that the inclusion of the nearest neighbors Coulomb repulsion
$V_{pd}$ is not necessary for binding as was concluded in several of
the exact diagonalization studies. For similar systems sizes as the previous 
simulations, but at zero temperature, we  looked at the pairing
correlation functions {\it as a function of distance} and found that
the pairing correlation functions decay quickly with distance and
are suppressed with increasing lattice size and $U_d$. The extended
s-wave channel is much more strongly suppressed than the $d_{x^2 -
y^2}$ channel. These trends are similar to those found in QMC
simulations of the one-band model.

The remainder of our report is organized as follows: in Section
\ref{HAMILTON} we define the Hamiltonian and the physical quantities
calculated and discuss the choice of model parameters. In Section
\ref{CPMC} we briefly describe the CPMC method, and then in Section
\ref{RESULTS} we present the results for both the binding energy for
holes and the pairing correlation functions. Finally in Section
\ref{CONCLUSIONS}, we discuss in detail our main conclusions.

\section{\label{HAMILTON} Three-band Hubbard Hamiltonian }

As proposed by Emery,\cite{emery} the three-band model physically mimics the
CuO$_2$ layer by having one Cu and two O atoms per unit cell, with the
Cu atoms arranged on a square lattice and the O atoms centered on the
edges of the square unit cells. In this layer, Emery assumed that the 
relevant orbitals are just those of copper $\  3d_{x^2-y^2}$ and 
oxygen $2p_x $ and $2p_y$. 

The Hamiltonian has the form:

\begin{eqnarray}
H & = &            \sum_{<j,k> \sigma} t_{pp}^{jk} 
              (p^{\dagger}_{j \sigma}p_{k\sigma} +
               p^{\dagger}_{k\sigma}p_{j \sigma} )  + 
         \epsilon_p \sum_{j\sigma} n^{p}_{j\sigma} +
         U_p\sum_j n_{j\uparrow}^p n_{j\downarrow}^p  \nonumber   \\
 &    &   \hspace{2cm}      + \epsilon_d \sum_{i\sigma} n^{d}_{i\sigma} + 
            U_d\sum_i n_{i\uparrow}^d n_{i\downarrow}^d   \\
 &    &    + V_{pd}\sum_{<i,j>} n_{i}^d n_{j}^p \nonumber +
 \sum_{<i,j>\sigma} t_{pd}^{ij} (d^{\dagger}_{i\sigma}p_{j\sigma} +
	                      p^{\dagger}_{j\sigma}d_{i\sigma})   \nonumber 
\end{eqnarray}
In writing the Hamiltonian, we adopted the convention that the operator
$d^\dagger_{i,\sigma}$ creates a {\it hole} at a Cu $3d_{x^2 - y^2}$
orbital and $p^\dagger_{j,\sigma}$ creates a {\it hole} in an O $2p_x$ or
$2p_y$ orbital. $U_d$ and $U_p$ are the Coulomb repulsions at the Cu and O
sites respectively, $\epsilon_d$ and $\epsilon_p$ are the
corresponding orbital energies, and $V_{pd}$ is the nearest neighbor
Coulomb repulsion. As written, the model has a Cu-O hybridization $
t_{pd}^{ij} = \pm t_{pd} $ with the minus sign occurring $ j = i +
\hat{x}/2 $ and $j = i - \hat{y}/2 $ and also hybridizaton $
t_{pp}^{jk} = \pm t_{pp} $ between oxygen sites with the minus sign
occurring for $ k = j - \hat{x}/2 - \hat{y}/2 $ and $ k = j +
\hat{x}/2 + \hat{y}/2 $. These phase conventions are illustrated in Fig.~1.

The system
 is said to be half-filled when there is one hole 
per Cu site.  At half-filling, for a wide range of parameters, its
ground-state is an anti-ferromagnetic insulator just like the one-band
Hubbard model. Its insulating gap , however, is a
charge transfer one in contrast with the gap in the the one-band model 
which is a Mott-Hubbard gap.

The values of the parameters in the Hamiltonian have been estimated by
a number of different constrained density functional and quantum
cluster calculations.\cite{hybertsen}${^-}$\cite{martin} Taken together,
these estimates define a reasonably limited range of the parameters
for the which the model might be labeled as ``physical.'' . We use 
Table~I of Ref 20, which summarizes results 
from density functional and cluster calculations, as the principal 
guideline for the parameter range we explored.

The Cu site Coulomb repulsion $U_d$ is large, making doubly occupancy
of Cu sites by two holes very unfavorable. The next largest parameter,
the charge transfer energy $\epsilon = \epsilon_p-\epsilon_d>0$, plays
a special role. Depending on the relative values of $\epsilon$, $U_d$ and
the bandwidth $W$ the system can be classified in different 
regimes .\cite{zaanen}
Estimates place the cuprate superconductors in the charge
transfer regime, in which $W < \epsilon < U_d$.
 The remaining parameters are relatively comparable
in magnitude. The role of the Cu-O hybridization $t_{pd}$ is also
special. This hybdization, through the super-exchange mechanism,
generates an anti-ferromagnetic exchange interaction between the spins in 
the Cu sites.

If $\epsilon \gg U_d $, the three-band model maps into a one-band
model with $t_{eff} \sim t_{pd}^2/ \epsilon$ and $U = U_d $. For $U_d
\gg t_{eff}$, the one-band model can in turn be mapped into the t-J model
with $J = 4t_{eff}^2/U_d$. Zhang and Rice \cite{zhang} have argued
that the t-J model can also be appropriate when $t_{pd} \ll \epsilon
,U_d,U_d - \epsilon$ and $\epsilon < U_d$. In real materials, 
$\epsilon/t_{pd}$ is estimated to
be $ \sim 2.7-3.7 $ .\cite{martin}  Therefore, besides  the
lack of conclusive evidence that the one-band model superconducts, it
is also unclear that the mapping among the most studied models is
appropriate for physical values of the parameters. In this paper we
will examine the three-band model in the estimated range of parameters to
detail the behavior of the model in its own right.

In what follows we scale all the energies by $t_{pd}$. 
 We take $V_{pd} = 0$ since it is small ($V_{pd} \le 0.8t_{pd}$ \cite{mahan}).
 This assumption simplifies the numerical
algorithm significantly and reduces the computational time by at least
a factor of four.

With the numerical method used, a variety of expectation values can be
computed. We focused on two key quantities: the binding energy for
doped holes and the pairing correlation functions as a function of the
distance.  The binding energy is defined as:
\begin{equation}
\Delta = (E_2 - E_0) - 2(E_1 - E_0)
\end{equation}
with $E_n $ the ground-state energy for n doped holes, where $E_0$
corresponds to the undoped (half-filled) case.  If negative, this
parameter indicates that it costs less to dope a second hole in the
vicinity of the first hole than it does to create two isolated
holes. In the thermodynamic limit $\Delta$ is expected to be the
binding energy of a Copper pair. We calculated $\Delta$ for systems
with $2 \times 2$, $4 \times 2$ and $6 \times 4$ unit cells. We are 
unaware of any previous
quantum Monte Carlo calculations of binding energy for the 
three-band model.

We also computed the extended s-wave and the $d_{x^2-y^2}$ pairing 
correlation functions:
\begin{equation}
P_\alpha(R) = < \Delta_\alpha^\dagger(R) \Delta_\alpha(0)>
\end{equation}
where
\begin{eqnarray}
& \Delta_\alpha(R) =  \sum\limits_{\vec{\delta}} f_\alpha(\vec{\delta}) \{ 
[d_{\vec{R}  \uparrow}d_{\vec{R}+\vec{\delta} \downarrow} -
d_{\vec{R}\downarrow}d_{\vec{R}+\vec{\delta}   \uparrow} ] &\\
&                                                            
+ [p^x_{\vec{R}  \uparrow}p^x_{\vec{R}+\vec{\delta} \downarrow} -
p^x_{\vec{R}\downarrow}p^x_{\vec{R}+\vec{\delta}   \uparrow} ]+
[p^y_{\vec{R}  \uparrow}p^y_{\vec{R}+\vec{\delta} \downarrow} -
p^y_{\vec{R}\downarrow}p^y_{\vec{R}+\vec{\delta}   \uparrow}] \} & 
\nonumber
\end{eqnarray}
with $\vec{\delta} = \pm \hat{x}, \pm \hat{y}$.  For extended s-wave
pairing $ f_{s^*}(\vec{\delta}) = 1 $ for all $\vec{\delta}$ and for
the $d_{x^2 - y^2}$ pairing, $f_{d}(\vec{\delta}) = 1 $ for
$\vec{\delta} = \pm \hat{x}$ and $f_{d}(\vec{\delta}) = -1 $ for
$\vec{\delta} = \pm \hat{y} $. Here $R$ denotes the distance between
unit cells (taken from copper ion to copper ion).  We will directly
report these correlations functions as a function of $R$. We will not
 report its $q=0$ spatial Fourier
transformation and partial sums like $S_\alpha(L)= \sum_{R\le L}
P_\alpha(R)$ as it has been done in previous works 
\cite{scalettar1}$^{-}$\cite{kuroki}, because the magnitude 
of these quantities are dominated
by a large peak in $P_\alpha(R)$ when $R$ is less than a few nearest
neighbor distances. Over these distances, $P_\alpha$ measures local
correlations among spin and charge fluctuations and has little
information about long-range pairing correlations.
In certain cases, we will also report the ``vertex contribution'' to  
the correlation functions (see, for example, White et. al \cite{white1}) 
defined as follows:
\begin{equation}
V_\alpha(R) =  P_\alpha(R) - \stackrel{-}{P}_\alpha(R)
\end{equation}
where $\stackrel{-}{P}_\alpha(R)$  
is the contribution of two dressed non-interacting propagators. For each term
in $ P_\alpha(R)$ of the form $< c^\dagger_\uparrow c_\uparrow 
c^\dagger_\downarrow c_\downarrow>$ , $\stackrel{-}{P}_\alpha(R)$ has a term 
like $< c^\dagger_\uparrow c_\uparrow><c^\dagger_\downarrow c_\downarrow>$.
We find that the conclusions remain the same no matter which quantity we look 
at.

\section{\label{CPMC} Constrained Path Quantum Monte Carlo technique}

Our numerical method is extensively described and benchmarked
elsewhere.\cite{shiwei1} Here we only discuss its basic
approximation.  In the CPMC method, the ground-state wave function
$|\Psi_0\rangle$ is projected from a known initial wave function
$|\Psi_T\rangle$ by a branching random walk in an over-complete space
of Slater determinants $|\phi\rangle$.  In such a space, we can write
$|\Psi_0\rangle = \sum_\phi \chi(\phi) |\phi\rangle$.  The random walk
produces an ensemble of $|\phi\rangle$, called random walkers, which
represent $|\Psi_0\rangle$ in the sense that their distribution is a
Monte Carlo sampling of $\chi(\phi)$, that is, a sampling of the
ground-state wave function.

To completely specify the ground-state wave function, only
determinants satisfying $\langle\Psi_0|\phi\rangle>0$ are needed
because $|\Psi_0\rangle$ resides in either of two degenerate halves of
the Slater determinant space, separated by a nodal plane ${\cal N}$
that is defined by $\langle\Psi_0|\phi\rangle=0$. 
The sign problem occurs because walkers
can cross ${\cal N}$ as their orbitals evolve continuously in the
random walk. Asymptotically they populate the two halves equally,
leading to an ensemble that has zero overlap with $|\Psi_0\rangle$.
If ${\cal N}$ were known, we would simply constrain the random walk to
one half of the space and obtain an exact solution of Schr\"odinger's
equation.  In the constrained-path QMC method, without 
{\it a priori\/} knowledge of ${\cal N}$, we use a
trial wave function $|\Psi_T\rangle$ and require
$\langle\Psi_T|\phi\rangle>0$.  The random walk again solves
Schr\"odinger's equation in determinant space, but under an
approximate boundary-condition.  This is what is called the
constrained-path approximation.

The ground-state energy computed by the CPMC method is an upper bound.
The quality of the calculation clearly depends on the trial wave
function $|\Psi_T\rangle$. Since the constraint only involves the
overall sign of its overlap with any determinant $|\phi\rangle$, it
seems reasonable to expect the results to show some insensitivity to
$|\Psi_T\rangle$.  Through extensive benchmarking on the Hubbard
model, it has been found that simple choices of this function can give
very good results.\cite{shiwei1} 

Besides as starting point and as a condition constraining a random
walker, we also use $|\Psi_T\rangle$ as an importance
function. Specifically we use $\langle\Psi_T|\phi\rangle$ to bias the
random walk into those parts of Slater determinant space that have a
large overlap with the trial state. For all three uses of
$|\Psi_T\rangle$, it clearly is advantageous to have $|\Psi_T\rangle$
approximate $|\Psi_0\rangle$ as closely as possible. Only in the
constraining of the path does $|\Psi_T\rangle \not= |\Psi_0\rangle$ generate
an approximation.

All the calculations reported here are done with periodic boundary conditions.
 Mostly, we study closed shell cases, for which the 
corresponding  free-electron wave function is non-degenerate and is
translationally invariant. In these cases, the free-electron wave function,
represented by a single Slater determinant, is used as the trial wave 
function $|\psi_T > $.
(The use of an unrestricted  Hartree-Fock wave function as $|\psi_T > $
produced no significant improvement in the results).

However, to calculate the binding energy for holes,
we need to study the half filled case and then the 1 and 2 hole doped
cases. In the systems considered ($2 \times 2$, $4 \times 2$ and $6 \times 4$ unit cells) the 2 hole 
doped case 
corresponds to a closed shell case. The one hole doped case is a closed 
shell minus one hole. This missing hole can be put in either of two degenerate
orbitals corresponding to $\vec{k_1} \equiv (0,\pi)$ or 
$\vec{k_2} \equiv (\pi,0)$.
The trial wave function, constructed by putting the hole in 
either of these orbitals or a linear combination of them, give equally good 
results so we pick an arbitrary linear combination of them. 
For the half-filled case, the free electron wave function is 4-fold 
degenerate: $c^\dagger_{k_1,\uparrow}c^\dagger_{k_1,\downarrow}|CS>$, 
            $c^\dagger_{k_2,\uparrow}c^\dagger_{k_2,\downarrow}|CS>$,
	     $c^\dagger_{k_1,\uparrow}c^\dagger_{k_2,\downarrow}|CS>$,
             $c^\dagger_{k_2,\uparrow}c^\dagger_{k_1,\downarrow}|CS>$, where
$|CS>$ is the closed shell state with two holes below half-filling.
In the $2 \times 2$ system the state $|CS>$ has 2 holes, in the $4 \times 2$ system, 
it has 6 holes and in the $6 \times 4$ system it has 22 holes.
If we just arbitrarily pick one of these states, the energy 
has a significant difference. For example, in the $2 \times 2$ system it is 
of the order of $10^{-2}$  from the exact 
diagonalization result. To accurately compute the 
binding energy, this difference is too big. Therefore, we used the following 
procedure: we diagonalize the interacting part of the Hamiltonian in this
degenerate subspace, and obtained 2 states with energy proportional to
$U_d$ and 2 states with zero energy. Of the 2 states with zero energy only one
of them is a singlet. We used this state. It is represented by two 
Slater determinants:
$(c^\dagger_{k_1,\uparrow}c^\dagger_{k_1,\downarrow} -
  c^\dagger_{k_2,\uparrow}c^\dagger_{k_2,\downarrow})|CS>/\sqrt{2}$.
In Table 1,  for a $2 \times 2$ system with $\epsilon_p =3$ and $U_p = t_{pp} = 0$, 
we compare the energies obtained using the CPMC with the one and two
Slater determinants trial wave function and energies obtained from
 exact diagonalization. We see that using two Slater determinants improves
the accuracy by an order of magnitude or more. The accuracy has become
better than the closed shell case.

In a  typical run for a large system we set the average number of walkers 
to 600 and the time step to 0.03.  We perform 3000 steps before we start 
taking measurements and we do the measurements in 40 blocks of 500 steps each
to ensure statistical independence.

\section{\label{RESULTS} Results}

Shell structures are characteristic of finite-sized systems of
electrons. In our simulations these effects are most clearly seen in
the values of the energy and chemical potential as a function of the hole
density $n = N_h/N$, with $N$ the number of unit cells. 
A shell structure is perhaps most easily illustrated by first
considering the non-interacting problem. There, holes are added to the
system at the same energy until all the degenerate states of a given
shell are occupied and paired, that is, until the shell is closed. As
an ``open'' shell is filled, the total energy of the system varies
linearly with the doping and within a shell the chemical potential
thus is a constant. The number of states in a shell is twice the order
of the point group relating the wave vectors of the degenerate
states. Between shells the chemical potential is discontinuous. This
``gap'' is a finite size effect.

In the interacting problem the energy and chemical potential shows a
similar behavior. The number of states in a shell and the shell
boundaries are the same as the non-interacting problem for the range of 
interactions we examined.  For a
$4\times 4$ system, a typical shell structure for the three-band model
as a function of $U_d$ is illustrated in Figs.~2a and 2b were we show
the ground state energy $E_o$ and the chemical potential $\mu$ as a 
function of the hole density for $\epsilon = 3$ , $t_{pp} = U_p = 0$. 
For our finite system we defined
\begin{equation}
\mu (N) = E_o(N) - E_o(N-1)
\end{equation} 
which is the discrete version of the usual definition 
$\mu = \partial E_o/\partial N_h$. The $U_d=0$
case is the non-interacting case.  Its shell structure is most evident
in the chemical potential, Fig.~2b. A comparison of the
non-interacting case with the interacting cases does show several
differences. At half-filling, $n=1$, the chemical potential in
the interacting cases shows a jump whereas the non-interacting case
does not. This jump is not a finite-size effect but is the
charge-transfer gap induced by the interaction. Within a shell the
chemical potential of the interacting cases are varying weakly but
approximately linearly with doping. This variation is similar to
the behavior of the one-band Hubbard model\cite{furukawa,shiwei3}, but there,
to an excellent approximation, within a shell the energy for the
interacting cases varies linearly with doping and hence within a shell
the chemical potential is a constant. In Fig.~2a, one sees that the
energy versus doping curve has a minimum whose location varies as
a function of $U_d$. As $U_d$ increases this location shifts from the
fully doped position towards the half-filled position.  The one-band
Hubbard model displays similar doping and interaction dependencies.
\cite{moreo,furukawa,shiwei3}

\subsection{\label{BINDEN} Binding Energy for Holes}

We calculated the binding energy $\Delta$ for holes for systems with
$2\times 2$, $4\times 2$, and $6\times 4$ unit cells. A value of
$\Delta < 0$ indicates hole binding whereas a value of $\Delta > 0$
implies hole repulsion. It is important to note that studying systems
larger than $2\times 2$ is essential for the doping $\delta$ to have
values similar to those for which the real materials superconduct: for
$2\times 2$ unit cells with one hole, $\delta = 0.25$, and with 2
holes, $\delta = 0.5$, values which are outside the physically
interesting region.  On the other hand, for one and two holes the
$4\times 2$ case has $\delta = 0.125$ and $\delta = 0.25 $ while
the $6\times 4$ case has $\delta = 0.042$ and $\delta = 0.083$. These
values of doping span the range observed in the real materials.

Shown in Fig. 3, to benchmark the accuracy of our results, is a
comparison of our binding energies with those of exact
diagonalization. Specifically, we plotted $\Delta$ as a function
of $U_d$ for $2\times 2$ systems with $U_p = t_{pp} = 0 $ and $\epsilon = 2 $
and 3 and found good agreement with exact diagonalization results,
especially for the $\epsilon = 2$ case. For $\epsilon = 3$ the
differences are more pronounced but are still only of the order of
$10^{-3}$. As a general trend, we slightly {\it underestimate} the
binding energy. This figure is representative of our systematic error
in the estimation of the binding energy.

In Fig. 4 we present results for $4\times 2$ systems as function of $\epsilon$
as function of two values of $U_d$.  For $U_d=1$ we find that $\Delta$
decreases monotonically with $\epsilon$ but for $U_d =
4$ it has a shallow minimum at $\epsilon \sim 2$. Such a minimum means
that there is a broad range of $\epsilon$ for which the binding energy
does not change significantly. The existence of such a minimum,
occurring at approximately the same value of $\epsilon$, was observed in
the exact diagonalization studies of Ogata and Shiba\cite{ogata} and
commented upon by Martin.\cite{martin} In general holes bind
provided $\epsilon$ is below some threshold value. In what follows, we
set $\epsilon = 3$ which is a value consistent with the values obtained by LDA
and cluster calculations \cite{martin} and is also near the optimal 
value for binding.

In Fig.~5, we compare $\Delta$ for different systems sizes as a
function of $U_d$.  We find that the binding energy tends to saturate
as $U_d$ is increased. The saturation is consistent with the exact 
diagonalization
studies of Ogata and Shiba\cite{ogata} and of Stephan et al.
\cite{stephan} The binding energy tends to increase as the system 
size is increased. 
Over the range of system sizes studied, the
dependence of $\Delta$ on system size was not systematic enough to
allow us to extrapolate to infinite system sizes. In general $\Delta
\sim 10^{-2}$.

We also studied the effect of $U_p$ and $t_{pp}$ on $\Delta$. In
Fig. 6 we plot $\Delta$ as a function of $U_d$ for $U_p =0$ and $U_p
=1$. In Fig.~6a, we see that in the $2\times 2$ system a small value
of $U_p$ destroys the binding as previously reported by Hirsch et
al.\cite{hirsch} However, in the larger $4\times 2$ system
(Fig.~6b), the holes still bind with $U_p =1 $, although the binding
energy is smaller, confirming a speculation of Hirsch et al. that the
negative effect of $U_p$ on the binding might decrease as the system
size increases. A similar effect is seen when we plot $\Delta$ versus
$t_{pp}$ for $2\times 2$ and $4\times 2$ systems with $U_d = 6$, 
$\epsilon = 3$ and $U_p = 0$ (Fig. 7).  The
ability to bind decreases nearly linearly with $t_{pp}$ but in the
larger systems it decreases more slowly, so that there is still
binding for $t_{pp} = 0.3$.  Therefore, the effect of the Coulomb
repulsion on the O sites and hopping between them in reducing the
binding is diminished for larger system sizes. In contrast to the
position emphasized by Hirsch et al.\cite{hirsch} and by
Stephan et al.,\cite{stephan} the physical presence of a non-zero
$t_{pp}$ and $U_p$, while reducing the ability to bind, does not
necessarily mean that an unphysically large Cu-O Coulomb repulsion
$V_{pd}$ is essential for hole binding. Their conclusions were
influenced by the small system sizes they were able to study.

While the binding of two holes in the physically relevant range of
parameters is encouraging for superconductivity, perhaps a more
central question now is what happens when a third hole is added. Do
the three holes cluster or does the third hole sit apart from the
pair?  We do not examine these questions directly. Instead, we
calculate the superconducting pairing correlations.

\subsection{\label{PAIRING} Pairing Correlation Functions}

For both the extended s-wave and the $d_{x^2 - y^2}$ channels, we
studied the pairing correlation functions as a function of the
distance for lattices up to $6 \times 6$ unit cells. All calculations
were done for closed shell cases. For the $4 \times 4$ systems, we put
18 holes ($\delta = 0.125$), for the $6 \times 6$ system, 42 holes
($\delta = 0.167$), and for $6\times 4$, 26 holes ($\delta = 0.083$).
The distance is measured between unit cells (from copper ion to copper ion)
\cite{explain}.

In Fig. 8 we present the pairing correlations for the extended s-wave
channel of a $6\times 6$ system with $\epsilon = 3$ and $t_{pp} = 0.3$
for different values of the interaction $U_d$. The inset magnifies the
large distance behavior. We see that the short-ranged correlations are
enhanced with increasing $U_d$, but decay very quickly when $R$ is
increased. For large $R$ they simply fluctuate around $0$.  For the
same parameters, the $d_{x^2 - y^2}$ channel is plotted in
Fig. 9. Again, there is enhancement in the short-ranged part. Then as
$R$ is increased, there is again a decay, in this case not to zero but to a
small positive value. However, as $U_d$ is increased, this value
becomes even smaller.  These results clarify the findings of previous
quantum Monte Carlo simulations. These previous simulations were
limited to relatively high temperatures because of the sign
problem. Scalettar and coworkers\cite{scalettar2} found an attractive
pairing interaction in both the extended s and the d$_{x^2-y^2}$
channels but could not predict which would dominate with the lowering
of temperature and increasing of lattice size. The results of Figs.~8
and 9 clearly indicate that the d$_{x^2-y^2}$ correlations dominate.
This dominance is in direct opposition to the findings of Dopf and
coworkers\cite{dopf1,dopf2} that the extended s wave channel dominates. By
looking at the the zero wave number component of the Fourier transform
of the pairing correlation function, these researchers only observed
how the short-ranged correlations behaved and overshadowed the behavior
of the longer-ranged correlations.

In several previous quantum Monte Carlo simulations various
researchers found it more convenient to study just the vertex
contribution to the pairing correlation function, as defined in
section \ref{HAMILTON}. In Fig.~10 we plot this contribution to the
d$_{x^2-y^2}$ correlation function just shown in Fig.~9. We see that
there is positive contribution which again is enhanced by $U_d$ at
short distances but suppressed at larger distances. Thus, as emphasized
by Zhang et al.,\cite{shiwei2} this function seems to contain no more
information about possible long-range order than the full pairing
correlation function.

Since we restricted ourselves to closed shell cases, it is difficult
to do a useful finite-size scaling of our results.  An attempt to show
the effect of increasing system size on the correlations is given in
Fig. 11 where we plot the d$_{x^2-y^2}$ vertex contribution for $4\times 4$
and $6\times 6$ systems with the same parameters, keeping in mind that
the fillings are not the same. At short distances the correlations are
enhanced as the lattice size is increased but for larger distances
there is a crossover, and in fact, the vertex contribution is smaller
in the $6\times 6$ system by almost an order of magnitude. A similar
significant reduction is also seen in the long-range parts of the full
correlation functions.

We also studied the effect of $\epsilon$ and $t_{pp}$ on the pairing
correlations.  One will recall that increasing $t_{pp}$ weakened pair
binding while the binding was optimal for $\epsilon$ around 2. In
Fig. 12 we plot the bare correlation functions for different values of
$\epsilon$ with $U_d=6$ and $t_{pp}=0.3$ for a $6\times 4$ system with
$\delta = 0.083$. We find that the long-range pairing correlations are
not optimal when the pair binding is optimal. In fact, values of
$\epsilon$ less than the optimal value enhance the longer range part
of the correlation functions. A much smaller effect is obtained by
changing the value of $t_{pp}$, as evident in Fig. 13, where the
parameters are $U_d = 6$, $\epsilon = 3$ and the size is $6\times 6$
unit cells. If anything, the larger value of $t_{pp}$ seems to enhance
the long-range correlations slightly.

It is interesting that in all cases presented we see a crossover from
what happens at short distances to what happens at long distances. We
emphasize that examining quantities that integrate, even partially, the pairing
correlation function over distances can produce misleading results.

\section{\label{CONCLUSIONS} Summary and Conclusions}

Using a newly developed quantum Monte Carlo method, the
constrained-path method, we simulated the 3-band Hubbard model to
search for a possible superconducting ground state. We focused
principally on the calculation of the binding energy of a pair of
holes and the extended $s$ and $d_{x^2-y^2}$ superconducting pairing
correlation functions.  For the pair binding energy, we found that
holes bind for a wide physically relevant range of parameters and that
the binding increased as the system size is increased.  We also found
that when we included a hopping $t_{pp}$ between oxygen sites, the
binding decreased roughly linearly with $t_{pp}$, but in a larger
system it decreased much more slowly. The Coulomb repulsion $U_p$ on
the oxygen sites was found to suppress the binding, but again for
larger systems its effect became less significant. The increased
tendency to bind as the system size is increased seems to diminish the
importance of an Cu-O Coulomb repulsion $V_{pd}$ for holes to bind as
suggested in several exact diagonalization
studies.\cite{hirsch,stephan} In general, hole binding occurs
over a broad range of $\epsilon$ values centered
roughly around two.  In this respect our simulations support the exact
diagonalization results of Ogata and Shiba.\cite{ogata}

We calculated the pairing correlation functions for the extended $s$
and the $d_{x^2 - y^2}$ channels as a function of the distance.  For
the extended s channel, we found that as the Coulomb interaction $U_d$
on the Cu sites is increased, the long-range part of the correlation
functions is suppressed and fluctuates around zero. For the $d_{x^2 -
y^2}$ channel, we found that the correlations decay rapidly with
distance towards a small positive value. However, this value becomes
smaller as the interaction $U_d$ or the system size is increased. The
same behavior is observed for the vertex contributions to these
correlation functions. Unequivocally, the $d_{x^2-y^2}$ pairing
correlations dominate the extended $s$ pairing correlations. This
finding is opposite to that previously emphasized by Dopf and
coworkers.\cite{dopf1,dopf2,dopf3}

We also studied the effect of the charge transfer energy $\epsilon$ on
the pairing correlation functions and found that smaller values of
$\epsilon$ produce larger correlations. Similarly, we looked at the
effect of the hopping $t_{pp}$ between oxygen sites and found that it
has a very small effect but that larger values of $t_{pp}$ are
slightly more favorable for pairing. These trends are
counter to those that enhance pair binding.

A systematic study of the size dependence of the binding energy and
pairing correlation functions was difficult because the way the doping
dependence of closed shell changed with changing system sizes. We did
present a comparison of the $d_{x^2-y^2}$ pairing correlation function
for a $4\times 4$ system and a $6\times 6$ system. Although the hole
doping was slightly different, a significant reduction of both the
vertex contribution and the full correlation function was observed in
the larger system. This result suggests that the long-range pairing
correlation found are unlikely to persist in the thermodynamic limit.

In general, we found similar trends for the three-band model that were
found for the one-band Hubbard model.\cite{shiwei2} The additional
degrees of freedom in the three-band model do not seem to enhance
superconductivity in an obvious way.  One could argue that a Cu-O
Coulomb repulsion $V_{pd}$ is needed; however, we found that $V_{pd}$
is not necessary to obtain hole binding and speculate that it is
unlikely that the addition of such a small parameter to the model
would change the picture dramatically. The question remains, if the
holes do bind, that is, a purely electronic mechanism induces an
attractive interaction between holes, but this attraction does not
lead to electron pairing, what does it, if anything, lead to. Does
it lead to phase separation or some novel phase, like stripes? This
question is currently under investigation.

\section*{Acknowledgments}

We are thankful to Tinka Gammel for providing  exact diagonalization results.
The C++ program used
for this work incorporated the {\it MatrixRef} matrix classes written by S.R.
White, available at http://hedrock.ps.uci.edu.
We acknowledge many helpful discussions with J. Carlson, Richard L. Martin 
and B. H. Brandow.

\begin{table}
\caption{
Comparison of the exact  ground-state energy with the CPMC result with a one 
and two Slater determinant
trial wave function for a $2 \times 2$ system. Parameters are 
$\epsilon = 3$ and  $t_{pp} = U_p = 0$. The use of two Slater determinants as 
described in the text improves the accuracy by an order of magnitude or more.
}

\vspace{1cm}

\begin{tabular}{ c  c  c c } 
\hspace{0.25cm} $U_d$\hspace{0.25cm} & \hspace{1cm}   CPMC 1SD \hspace{1cm}  & \hspace{1cm} CPMC 2SD \hspace{1cm}  &  \hspace{1cm} Exact\hspace{1cm} \\ \hline
1     &  -5.0613(3) &  -5.0764(2)  &   -5.076977   \\ 
2     &  -4.8475(9) &  -4.8789(7)  &   -4.880047   \\ 
4     &  -4.6073(9) &  -4.6615(6)  &   -4.661723   \\ 
6     &  -4.4884(9) &  -4.5468(6)  &   -4.547436   \\  
\end{tabular}

\label{table1}
\end{table}

\begin{figure}

\caption{
Phase convention for the hopping matrix elements. The copper $d_{x^2-y^2}$ 
orbital is surrounded by the oxygen $p_x$ and $p_y$ orbitals. The hopping
matrix elements are shown with their corresponding phase.
}
\label{fig1}

\vspace*{0.9cm}

\caption{(a) Ground state energy per unit cell as a function of the 
hole  density for a $4 \times 4$
system with $\epsilon = 3$ , $t_{pp} = U_p = 0$. The energy decreases 
monotonically
for $ U_d=0$  but goes through a minimum for $U_d > 0$. For large $U_d$ the 
minimum appears at half-filling.(b) Chemical potential as a function of 
the hole  density for the same parameters. The shell structure is evident. 
For large $U_d$ a charge transfer gap appears at half-filling.
}
\label{fig2}
\vspace*{0.9cm}

\caption{
Comparison of the binding energy as a function of $U_d$ with exact 
diagonalization results for $2 \times 2$ systems with $t_{pp} = U_p = 0$ and
$\epsilon =2$ and $3$. Our
results are in reasonable good agreement with the exact ones.
}
\label{fig3}
\vspace*{0.9cm}

\caption{
Binding energy $\Delta$ as a function of $\epsilon$ for a $4 \times 2$ system
with $t_{pp} = U_p = 0$. 
For $U_d=1$ it decreases 
monotonically but for $U_d =4$ it has a shallow minimum near $\epsilon \sim 2$.
}
\label{fig4}

\vspace*{0.9cm}

\caption{
Binding energy $\Delta$ as a function of $U_d$ for different system sizes 
with $\epsilon = 3$ and $t_{pp} = U_p = 0$. The binding
increases as the system size is increased.
}
\label{fig5}

\vspace*{0.9cm}

\caption{
Effect of $U_p$ on the binding energy, $\epsilon = 3$ and $t_{pp} = 0$. 
 (a) In the $2 \times 2$ system the binding  
disappears completely for $U_p =1$. (b) In the $4 \times 2$ system 
the binding is still present for $U_p =1 $ although is slightly suppressed.
}
\label{fig6}

\vspace*{0.9cm}

\caption{
Effect of $t_{pp}$ on the binding energy for  $2 \times 2$ and 
$4 \times 2$ systems with 
$\epsilon = 3$,   $U_d =6$ and $U_p = 0$. The 
binding energy decreases roughly linearly with $t_{pp}$ but it decreases
much more slowly in the larger system.
}
\label{fig7}

\vspace*{0.9cm}

\caption{
Extended s-wave pairing correlation functions vs. distance for different
values of $U_d$ for a $6 \times 6$ system with 
$\epsilon = 3$ and $t_{pp}=0.3$. The correlations
decay rapidly and oscillate around zero at larger distances.
}
\label{fig8}

\vspace*{0.9cm}

\caption{
d-wave pairing correlation functions vs. distance for different values of
$U_d$ with the same parameters as in Fig. 8. The correlations decay
towards a small positive value at large distances.
}
\label{fig9}

\vspace*{0.9cm}

\caption{
Vertex contribution for the d-wave pairing correlation functions vs. 
distance for the same parameters as in Fig. 8. It converges to a small value 
at large distances and this value decreases as $U_d$ increases.
}
\label{fig10}

\vspace*{0.9cm}

\caption{
Comparison of the vertex contribution for the d-wave pairing correlations
for two different system sizes with $U_d = 6$, $\epsilon = 3 $ 
and $t_{pp} = 0.3$.
Although the fillings are slightly different in each case, a strong reduction 
is seen in the larger system.
}
\label{fig11}

\vspace*{0.9cm}

\caption{
d-wave pairing correlation functions for different values of $\epsilon$ with
$U_d=6$, $U_p = t_{pp} = 0$ for a $6 \times 4$ system 
with $\delta = 0.083$. The smaller
values of $\epsilon$ are more favorable for pairing.
}
\label{fig12}

\vspace*{0.9cm}

\caption{
Effect of $t_{pp}$ on the d-wave pairing correlation functions in a 
$6 \times 6$ system with $U_d=6$ and $\epsilon = 3$. A very small 
increase is seen as $t_{pp}$ is increased.
}
\label{fig13}

\end{figure}


\begin{thebibliography}{99}
\bibitem{explain} There are cases where for a same distance R between unit 
cells but different orientations of the vector $\vec{R}$, the value of the 
correlation function is different. This occurs either due to the geometry
of the unit cell or the fact that we used a rectangular lattice. In these cases
we averaged the different values of the correlation functions.
\end{thebibliography}
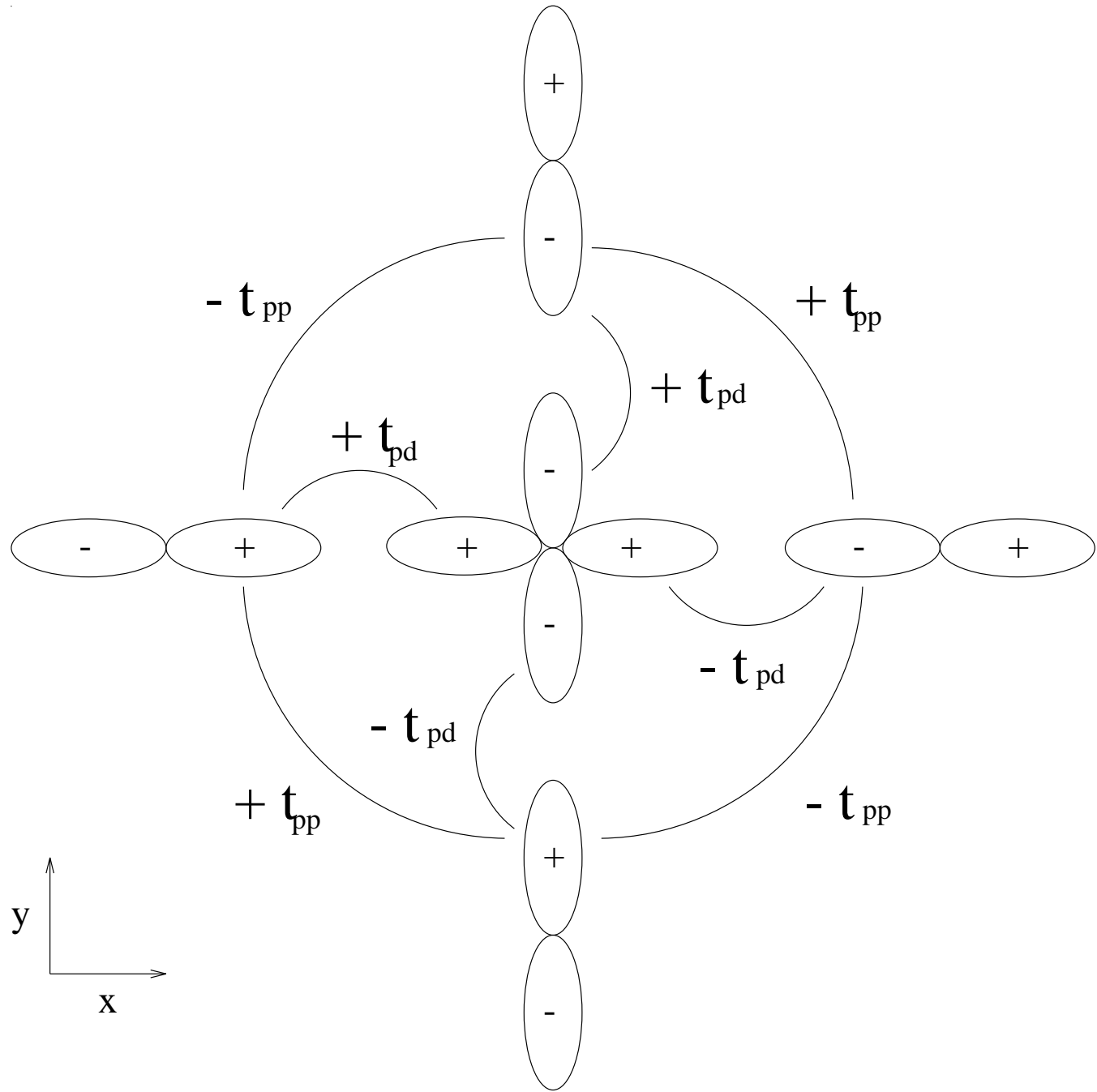
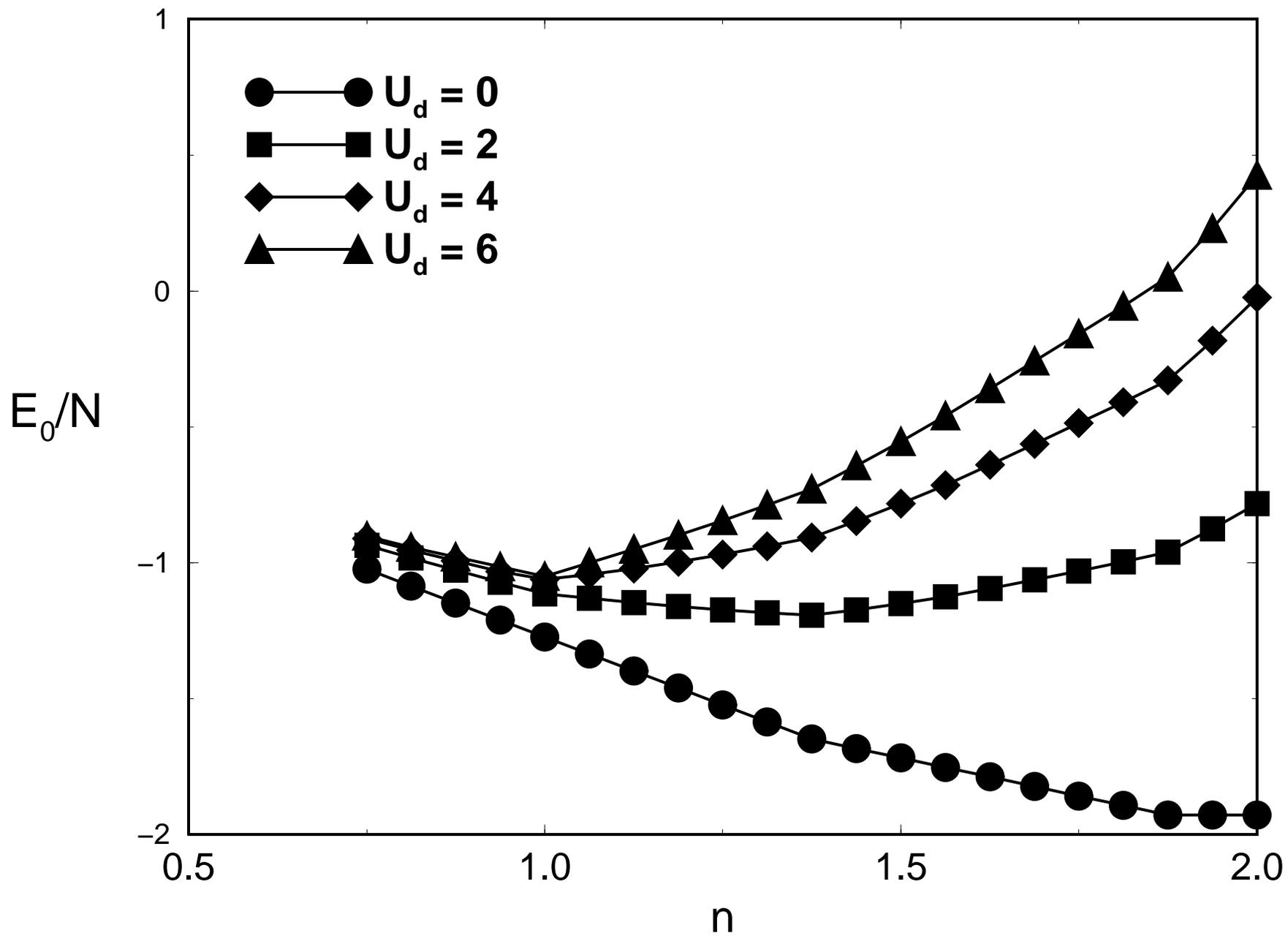
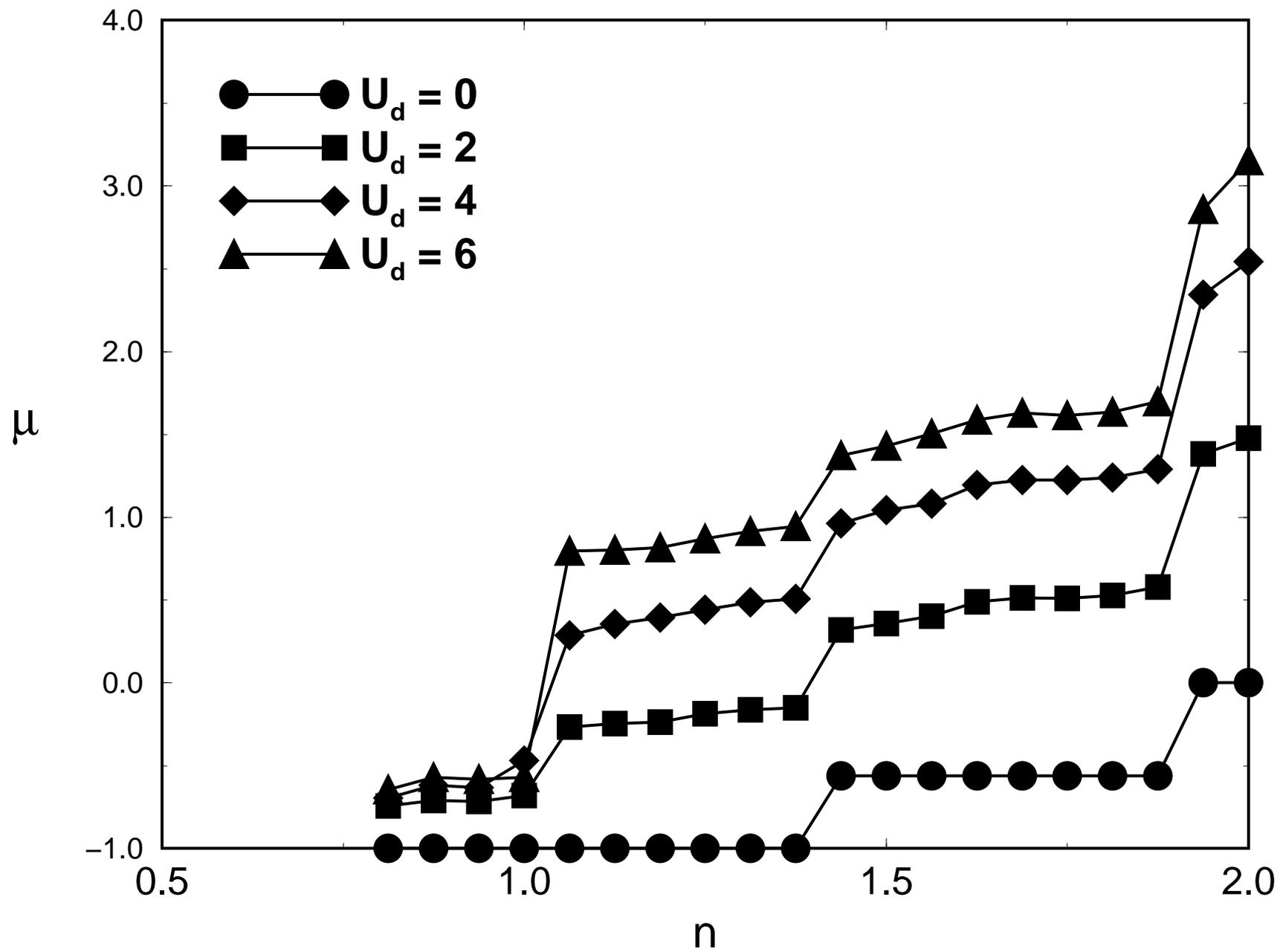
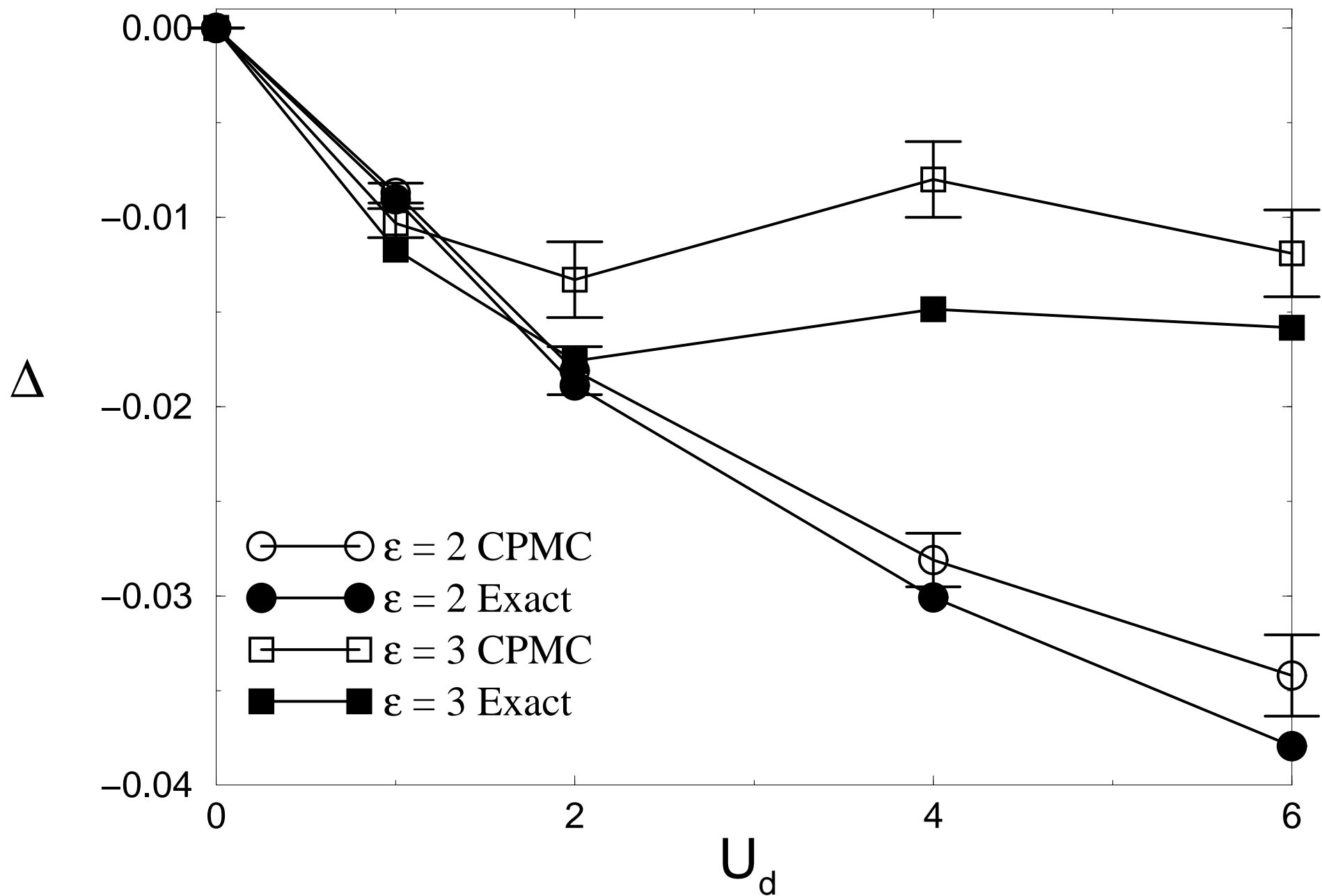
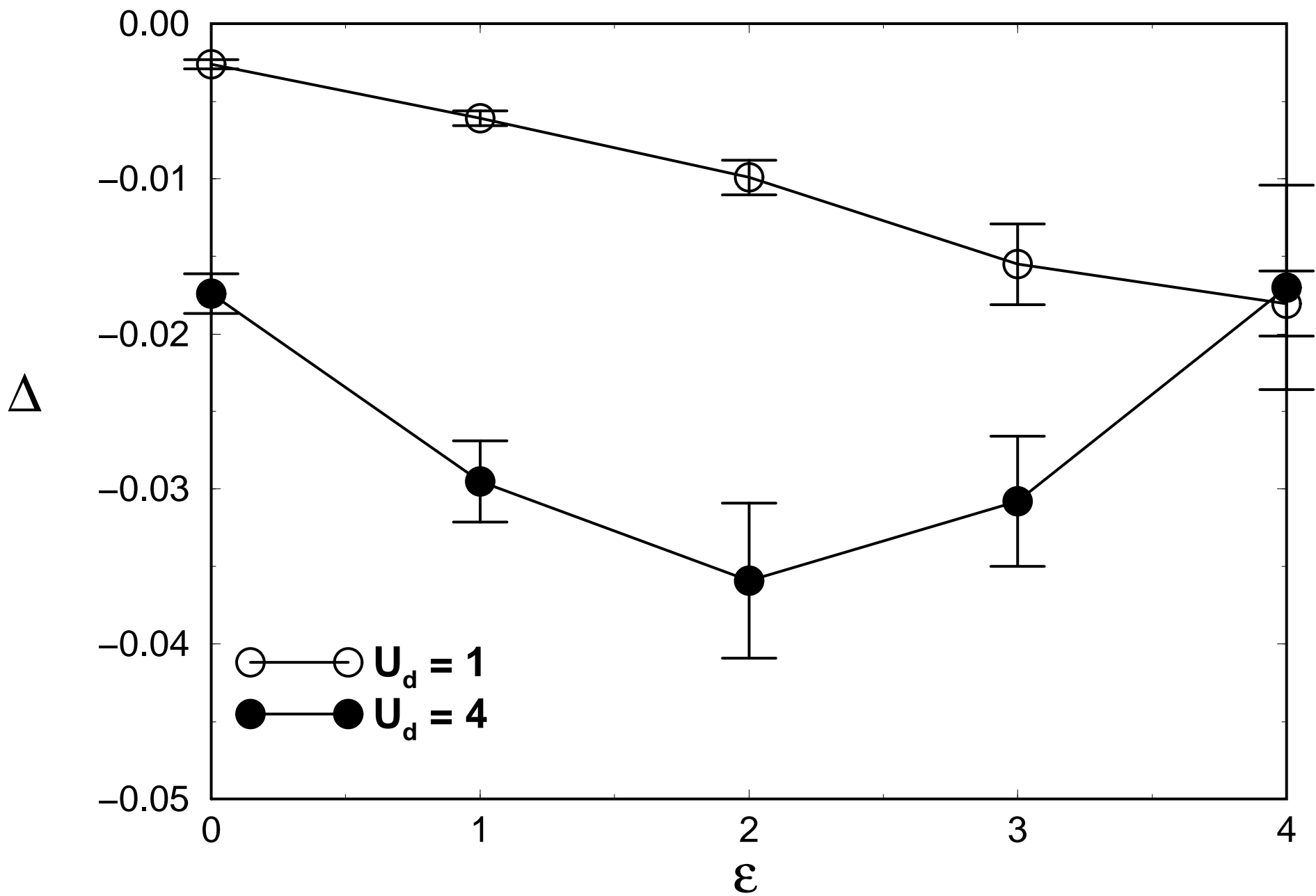
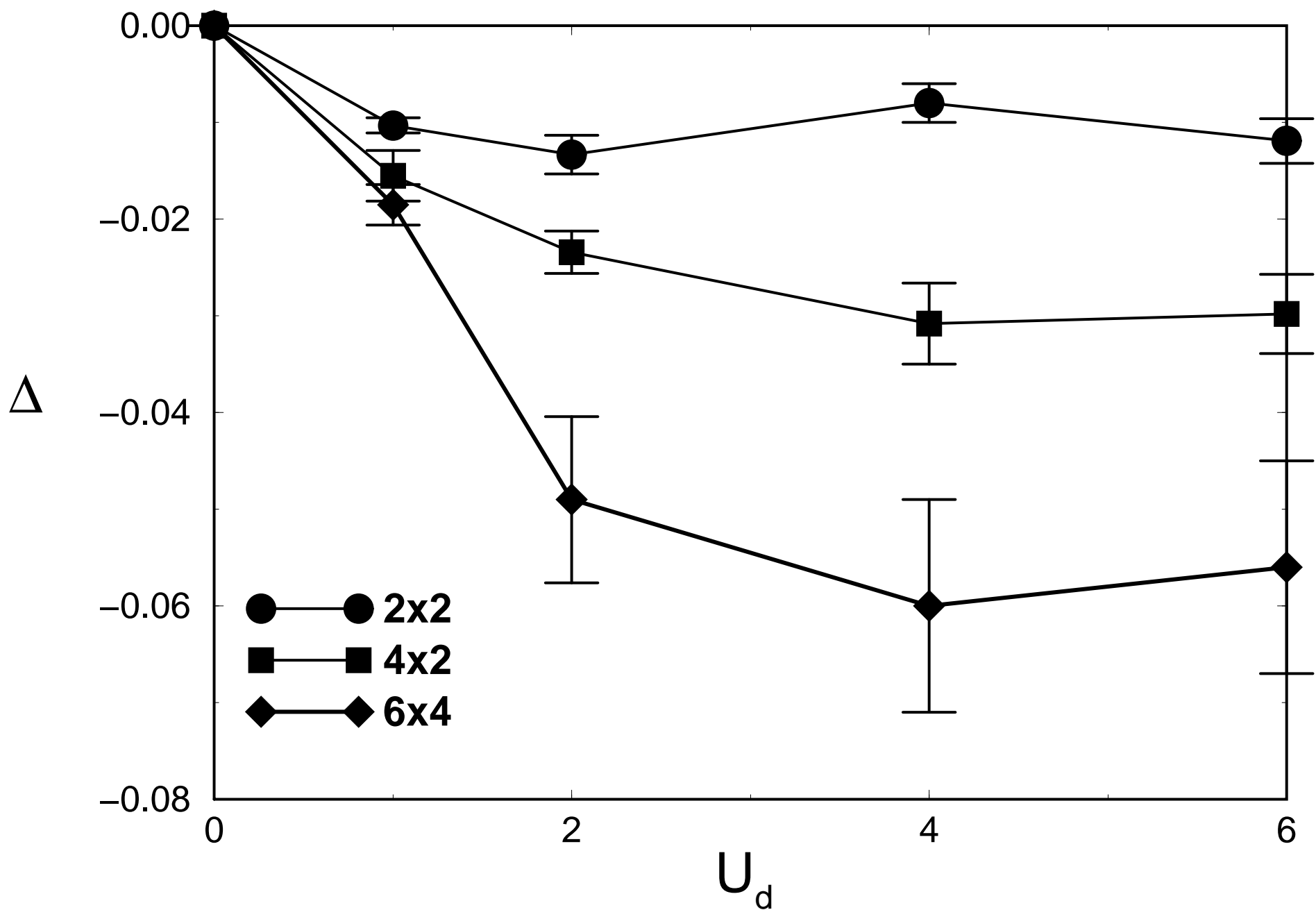
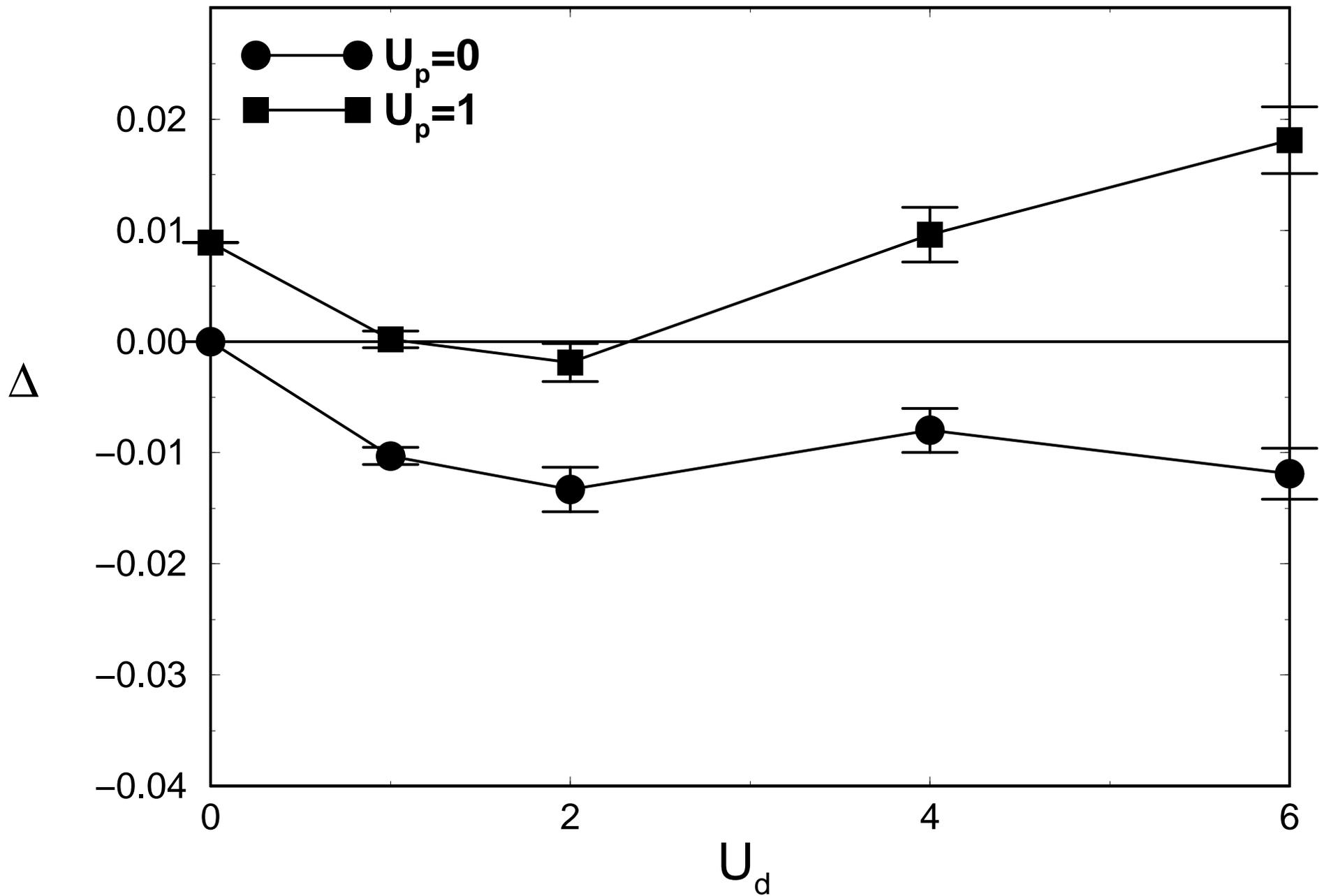
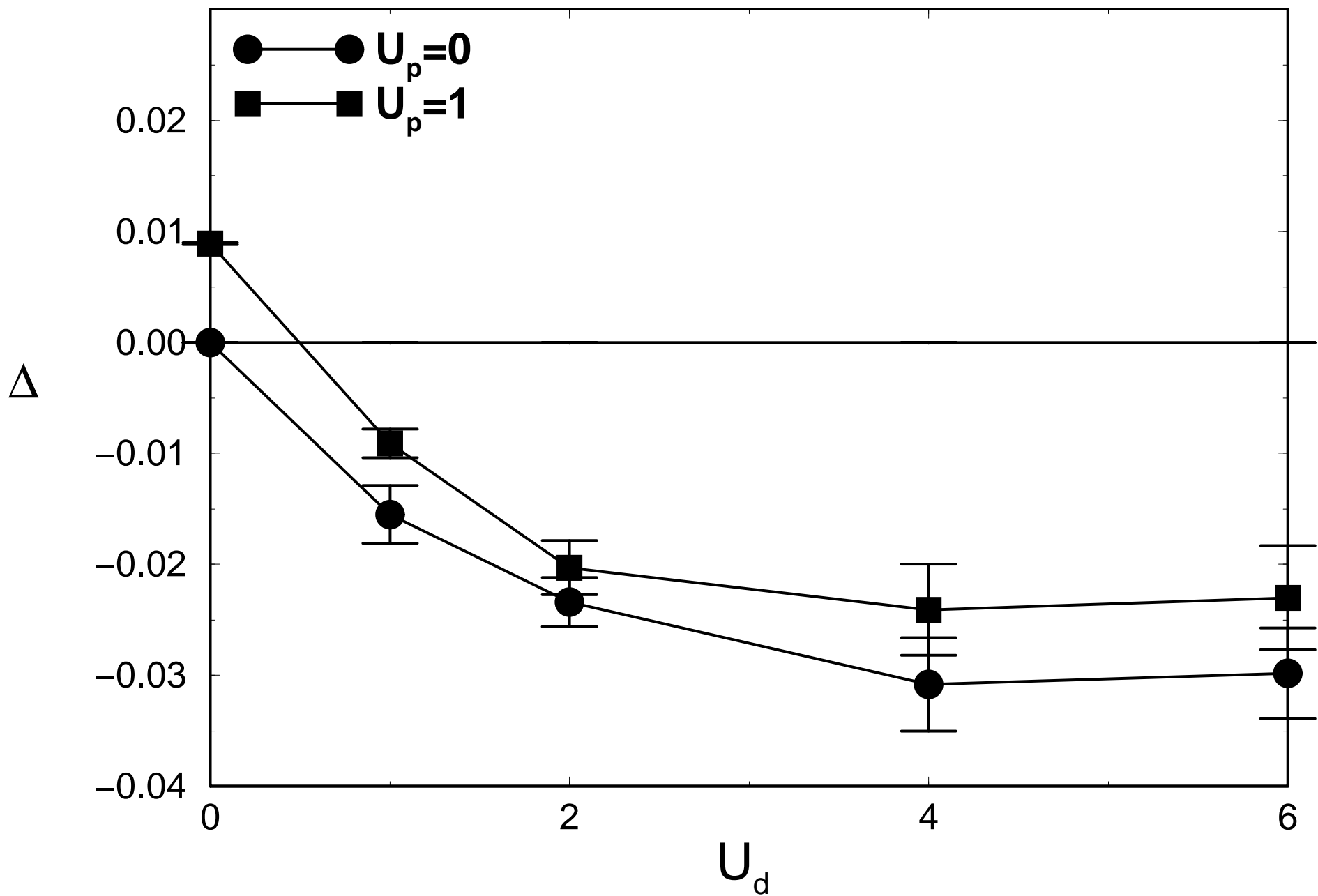
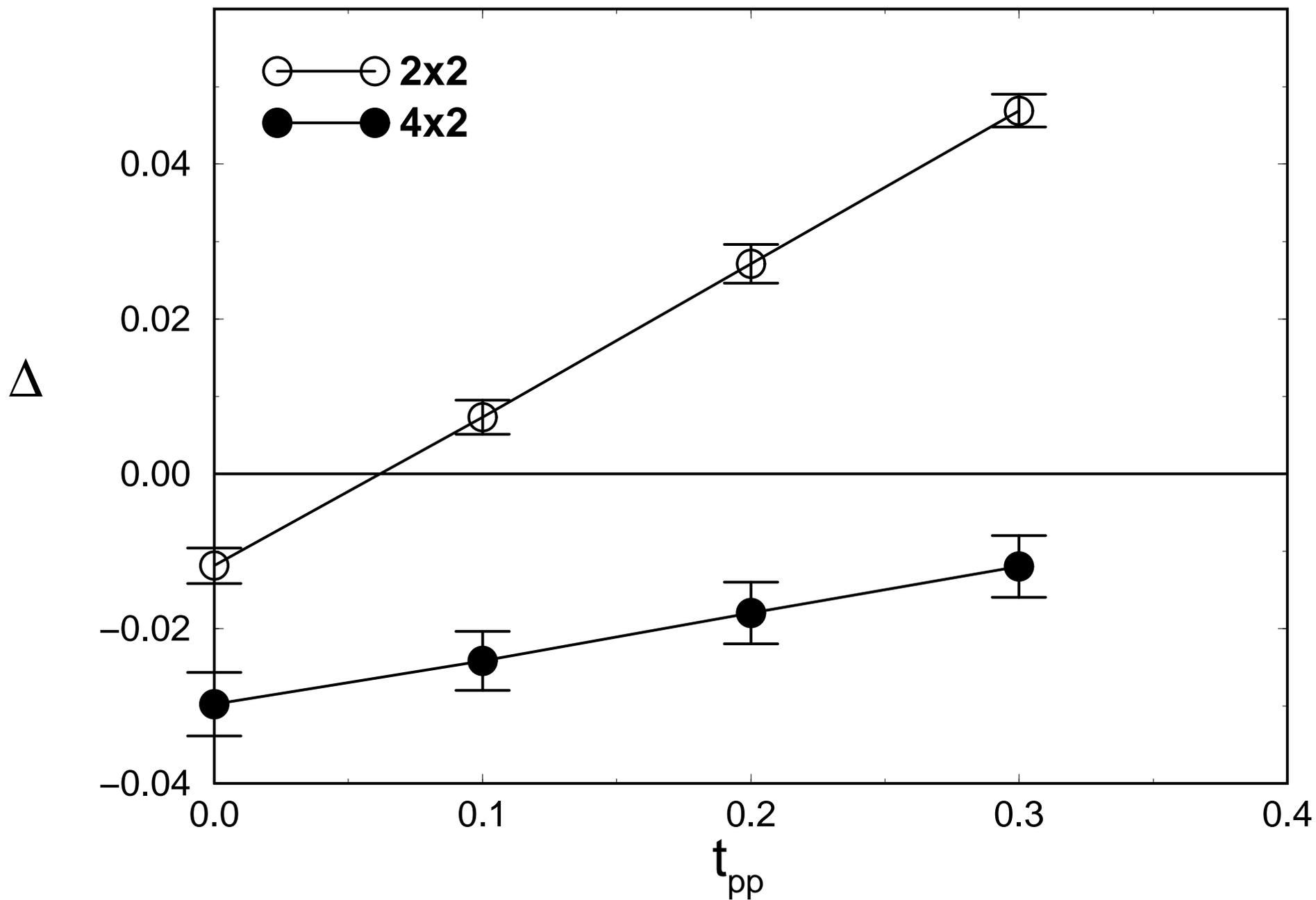
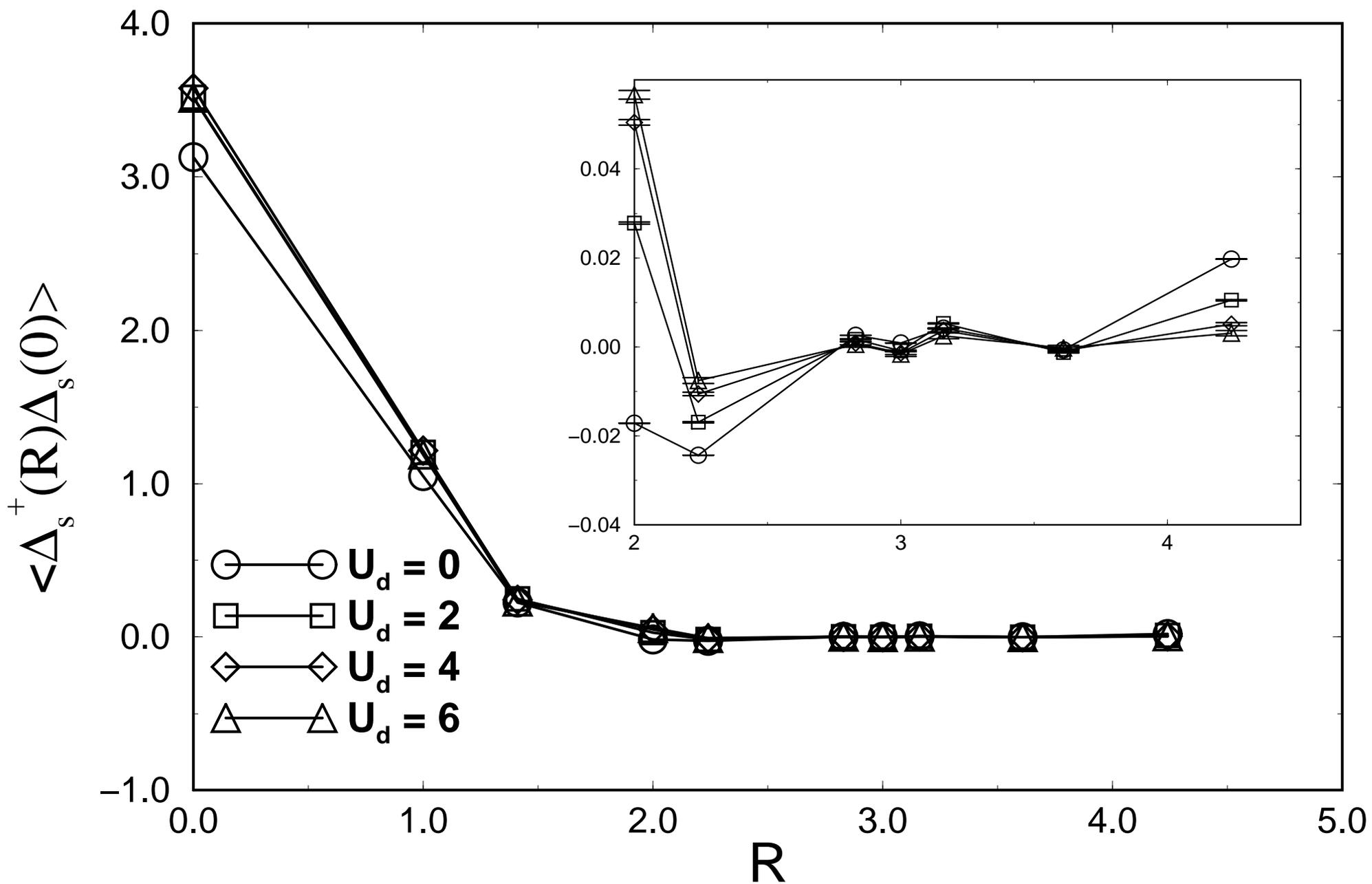
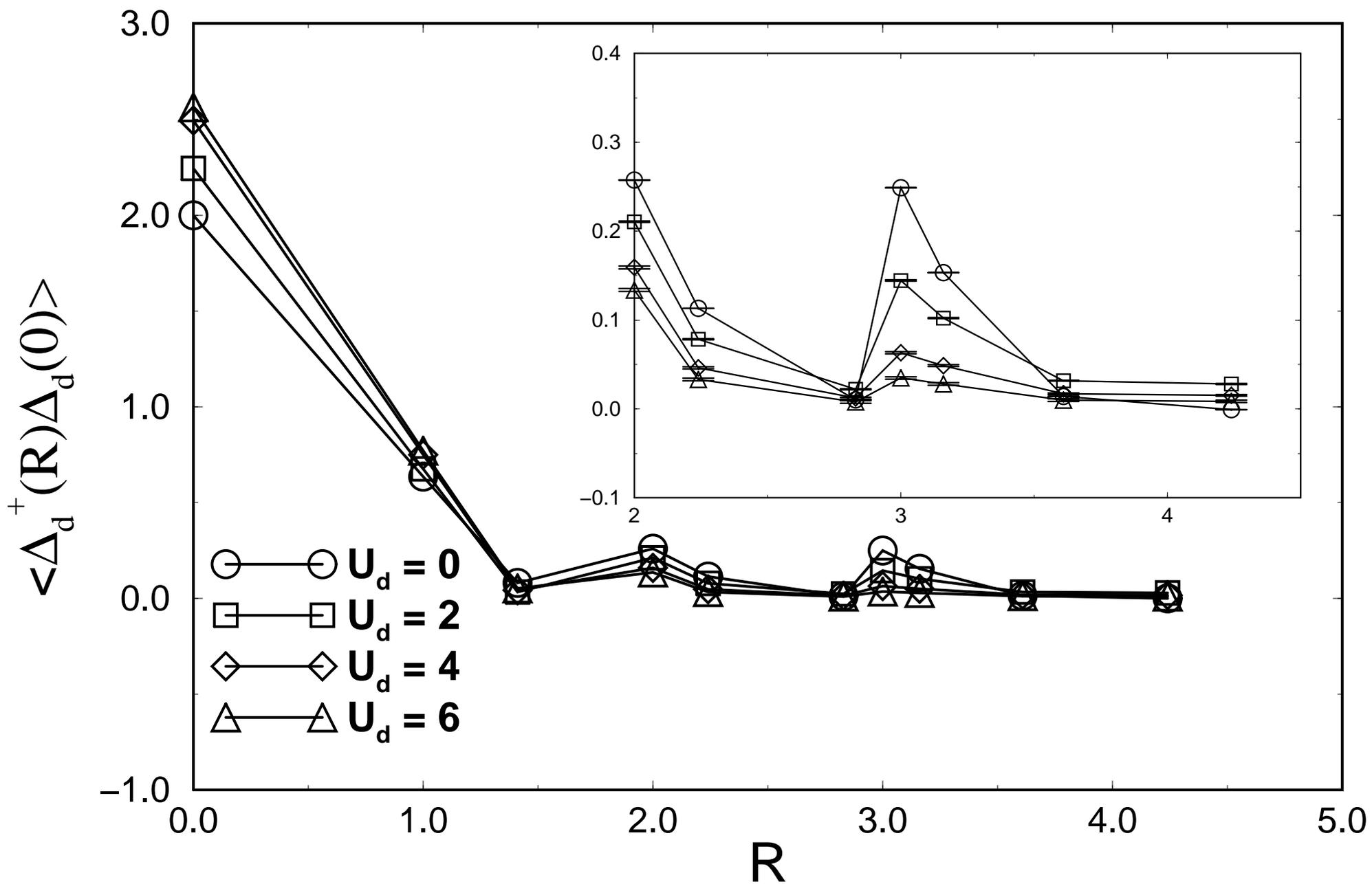
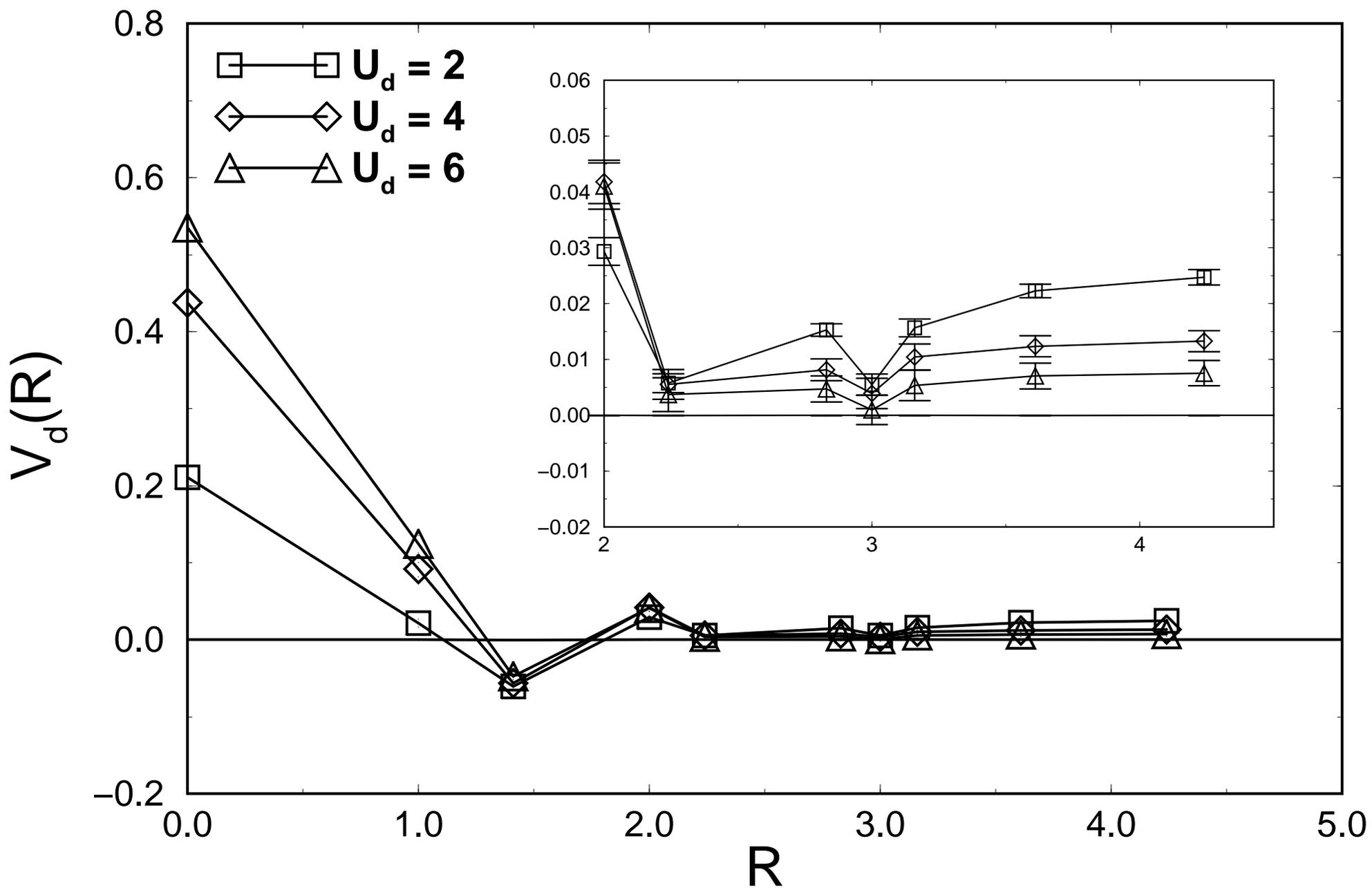
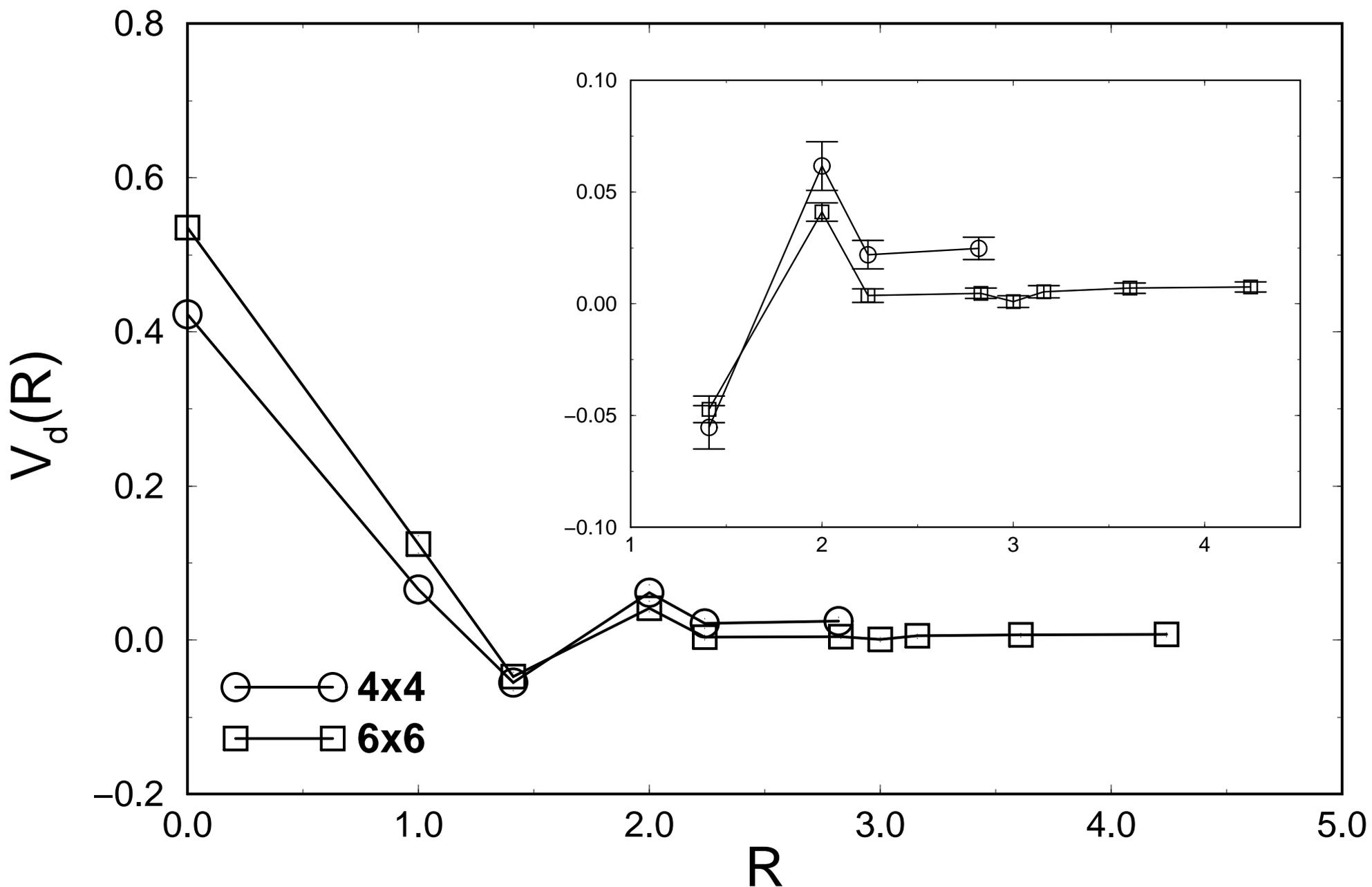
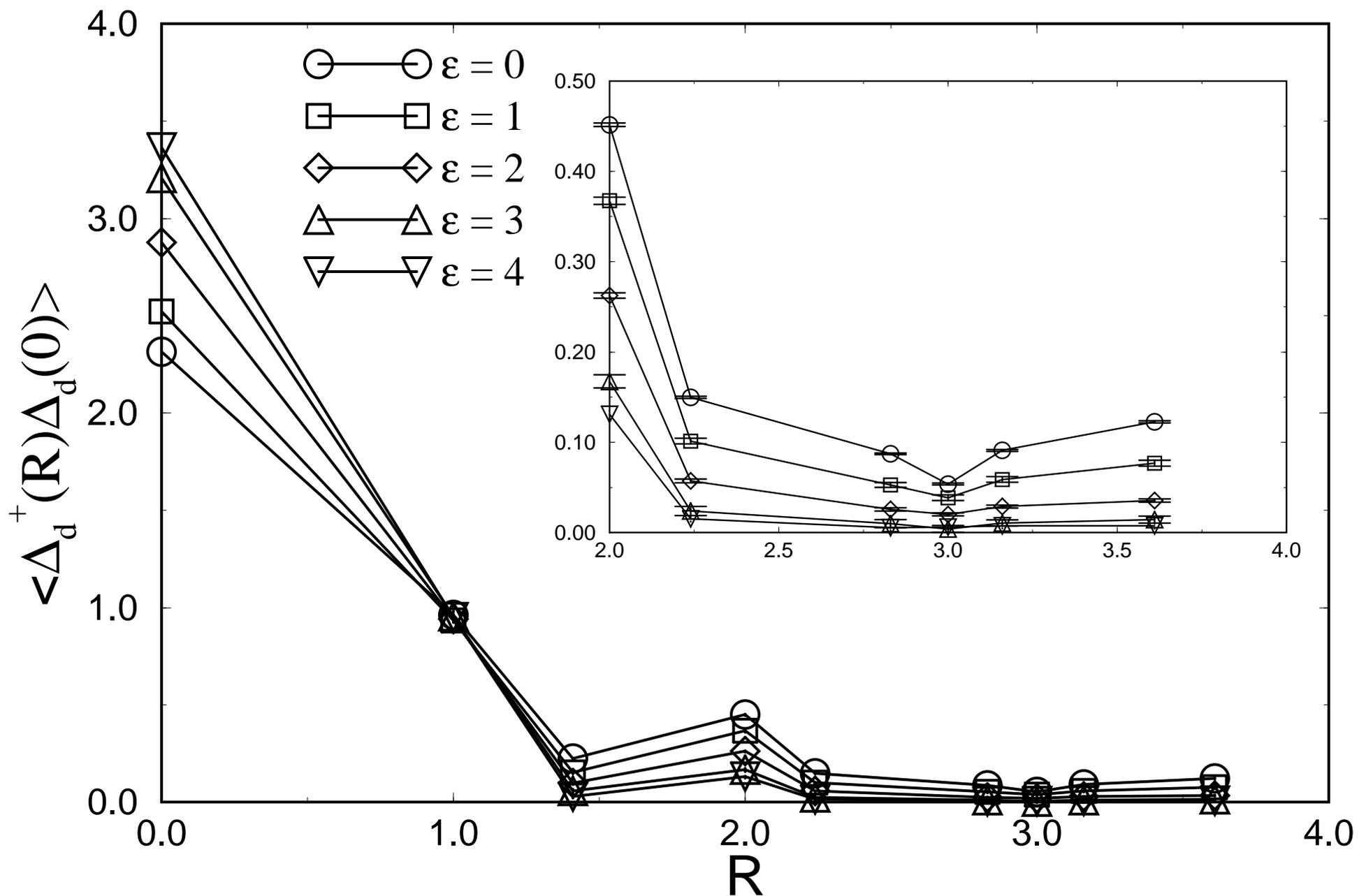
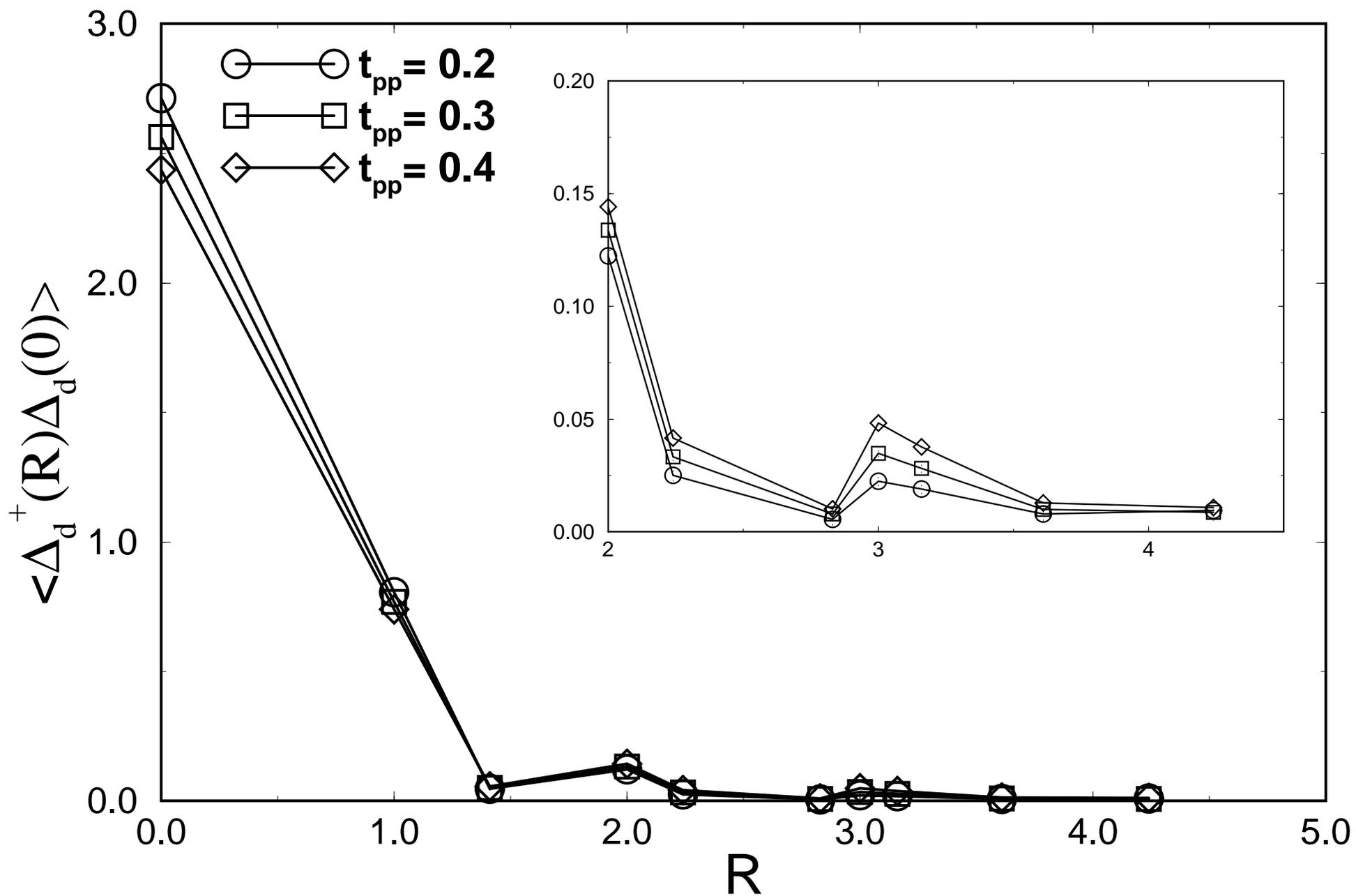
\end{document}